\documentclass[10pt,letterpaper,twocolumn]{article} 

\usepackage{ol2}
\usepackage[draft]{hyperref}
\usepackage{amsmath}

\begin{document}

\twocolumn[ 

\title{Topological Anderson phase in quasi-periodic waveguide lattices}


\author{Stefano Longhi}
\address{Dipartimento di Fisica, Politecnico di Milano and Istituto di Fotonica e Nanotecnologie del Consiglio Nazionale delle Ricerche, Piazza L. da Vinci 32, I-20133 Milano, Italy (stefano.longhi@polimi.it)}
\address{IFISC (UIB-CSIC), Instituto de Fisica Interdisciplinar y Sistemas Complejos, E-07122 Palma de Mallorca, Spain}

\begin{abstract}
The topological trivial band of a lattice can be driven into a topological phase by disorder in the system. This so-called
  topological Anderson phase has been predicted and observed for uncorrelated static disorder, while in the presence of 
correlated disorder conflicting results are found. Here we consider a Su-Schrieffer-Heeger (SSH) waveguide lattice in the trivial topological phase, and 
show that quasi-periodic disorder in the coupling constants can drive the lattice into a topological non-trivial phase. 
A method to detect the emergence of the  topological Anderson phase, based on 
light dynamics at the edge of a quasi-periodic waveguide lattice, is suggested.
\end{abstract}

 ] 


{\it Introduction.} Waveguide lattices provide a useful platform in integrated photonics to explore a wide variety of phenomena typical of crystalline materials \cite{r1,r2,r3,r4}, such as 
Anderson localization and topological transport \cite{r5,r6,r7,r8,r9,r10,r11}.  While strong disorder generally drives a topological non-trivial phase of a lattice band into a trivial one owing to Anderson localization, it came as a surprise the discovery that the reverse phenomenon can happen \cite{r12}, i.e. that static disorder can drive a topological trivial band into a nontrivial one 
\cite{r12,r13,r14,r16,r17,r18,r19}. Such a disorder-driven topological phase, referred to as  topological Anderson phase, has been recently observed in a two-dimensional photonic Floquet topological insulator \cite{r20}  and in disordered atomic wires \cite{r21} implementing the famous  Su-Schrieffer-Heeger (SSH) model of polyacetylene. Currently, it is a matter of debate whether different kinds of disorder can prevent the appearance of the  topological Anderson phase \cite{r23,r24,r25}.
Correlated disorder and bond disorder seem to prevent certain two-dimensional systems to enter into the  topological Anderson phase \cite{r23,r24}, while in other models  topological Anderson phase persists under incommensurate disorder \cite{r25}. 
While a precise control over disorder is challenging in real materials, photonic waveguide lattices can offer a suitable platform to control bond disorder via precise engineering of coupling constants between adjacent waveguides \cite{r26,r27}, and thus to investigate the impact of controlled disorder on topological Anderson phase.\\
\\
In this Letter we consider  topological Anderson phase in a one-dimensional SSH waveguide lattice  \cite{r17,r21} with incommensurate disorder in the inter-dimer bonds, and suggest an experimentally feasible spectral method
to detect the phase transition. It is proven that a topological Anderson phase can be induced by quasiperiodic disorder, and that edge dynamics measurements can detect the topological phase transition despite disorder-induced localization.\\
\\
{\it The SSH lattice with incommensurate disorder.} 
We consider an array comprising $2N$  optical waveguides with tailored coupling constants, as schematically shown in Fig.1. For alternating spacing between guides, this lattice realizes the SSH model, which is known to show the  topological Anderson phase transition under uncorrected disorder \cite{r17,r21}. In a disorder-free lattice with intra- and inter-dimer hopping amplitudes $t_1$ and $t_2$, the SSH lattice exhibits  two  topologically  distinct phases, characterized by a distinct winding number $Q=0$ for $t_1>t_2$ and $Q=1$ for $t_1<t_2$, separated by a phase  transition at the gap closing point $t_1=t_2$ \cite{r17,r21,r28,r29}. Topological zero-energy edge states exist in the non-trivial topological phase $t_1<t_2$, and they are robust against weak-to-moderate disorder in the hopping amplitudes \cite{r30}. Here we consider a SSH chain in the topological trivial phase $t_1>t_2$ with uniform intra-dimer hopping $t_1$ but with incommensurate disorder $V_n=V \cos (2 \pi \alpha n)$ added to the inter-dimer hopping $t_2$,  where $\alpha$ is irrational.  Light propagation in the waveguide array is described by coupled-mode equations \cite{r1,r2,r3,r29}
\begin{eqnarray}
i \frac{da_n}{dz} & = & t_1 b_n+(t_2+V_{n-1}) b_{n-1} \\
i \frac{db_n}{dz} & = & t_1 a_n+(t_2+V_n) a_{n+1} 
\end{eqnarray}
 \begin{figure}[htb]
\centerline{\includegraphics[width=8.4cm]{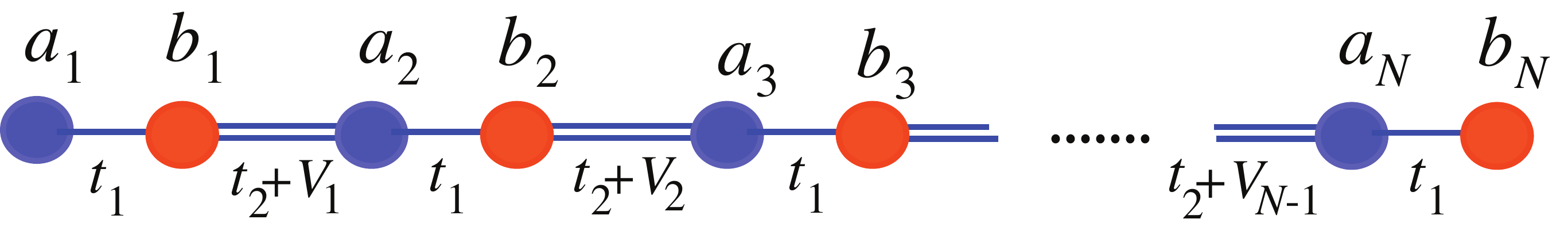}} \caption{ 
(Color online) Schematic of a dimeric SSH waveguide lattice comprising $N$ unit cells with uniform intra-dimer hoppping amplitude $t_1$ and with incommensurate disorder of the inter-dimer hopping amplitude $t_2+V_n$, with $V_n=V \cos (2 \pi \alpha n)$ and $\alpha=(\sqrt{5}-1)/2$.}
\end{figure} 
with $b_{0}=a_{N+1}=0$ for open boundary conditions (OBC). We assume $\alpha=(\sqrt{5}-1)/2$, which is approximated by the sequence $\alpha \simeq q_{n-1}/q_n$ of ratios of Fibonacci numbers $q_n$, defined by the recursive relation $q_0=0$, $q_1=1$, $q_{n+1}=q_{n}+q_{n-1}$. 
A typical energy spectrum $E=E_l$ of the SSH lattice versus disorder strength $V/t_1$ is shown in Fig.2(a) for a chain comprising $N=q_{11}=987$ unit cells and for $t_2/t_1=0.3$. Note that, owing to chiral symmetry, the energy spectrum is symmetric at around $E=0$. The localization properties of the $l$-th eigenmode $(a_n^{(l)}, b_n^{(l)})$ is determined by the inverse of the participation ratio (IPR) $I_l$, given by \cite{r27,r30}
\[
I_l= \frac{\sum_n \left( |a_n^{(l)}|^4+  |b_n^{(l)}|^4\right) }{\left( \sum_n (|a_n^{(l)}|^2+|b_n^{(l)}|^2) \right)^2}.
\]
Note that $I_l$ is bounded between zero and one, with $I_l \simeq 0$ for a delocalized state and $I_l \sim 1$ for a fully localized state at one site. Figure 2(b) shows the behavior of $I_l$, for all $2N$ modes, versus $V/t_1$. The figure indicates the existence of three 
regions as the disorder strength $V/t_1$ is increased: (i) Region I, defined by the range $0<V<t_2$, where most of the modes remain delocalized; (ii) Region II, defined by the range $t_2<V<2t_1$, where most of the eigenstates become localized; (iii) Region III, defined by the interval $V>2t_1$, where all eigenstates are localized. Note that in regions I and II the main (central) energy gap remains open, it closes at $V=2t_1$, above which two nearly-degenerate zero-energy modes emerge. The transition from region I to region II is related to the bulk properties of the quasi-periodic lattice. In the disorder-free limit $V=0$, the lattice shows two energy bands separated by a gap of width $2(t_1-t_2)$. For a nonvanishing disorder strength $V$, the bulk energy spectrum can be computed by considering the rational approximant $\alpha=q_{n-1}/q_n$ in the large $n$ limit, i.e by considering a superlattice comprising $2q_n$ sites in its unit cell \cite{r31}. In this case, the energy spectrum is composed by $2q_n$ narrow bands separated by gaps. Numerical analysis indicates that the spectral measure of the absolutely continuous spectrum \cite{r31}, defined as the sums of the widths of all the $2q_n$ energy bands, vanishes as $V \rightarrow t_2$, indicating that Bloch states become more and more localized as the boundary between regions I and II is approached.\\
\\
 \begin{figure}[htb]
\centerline{\includegraphics[width=8.7cm]{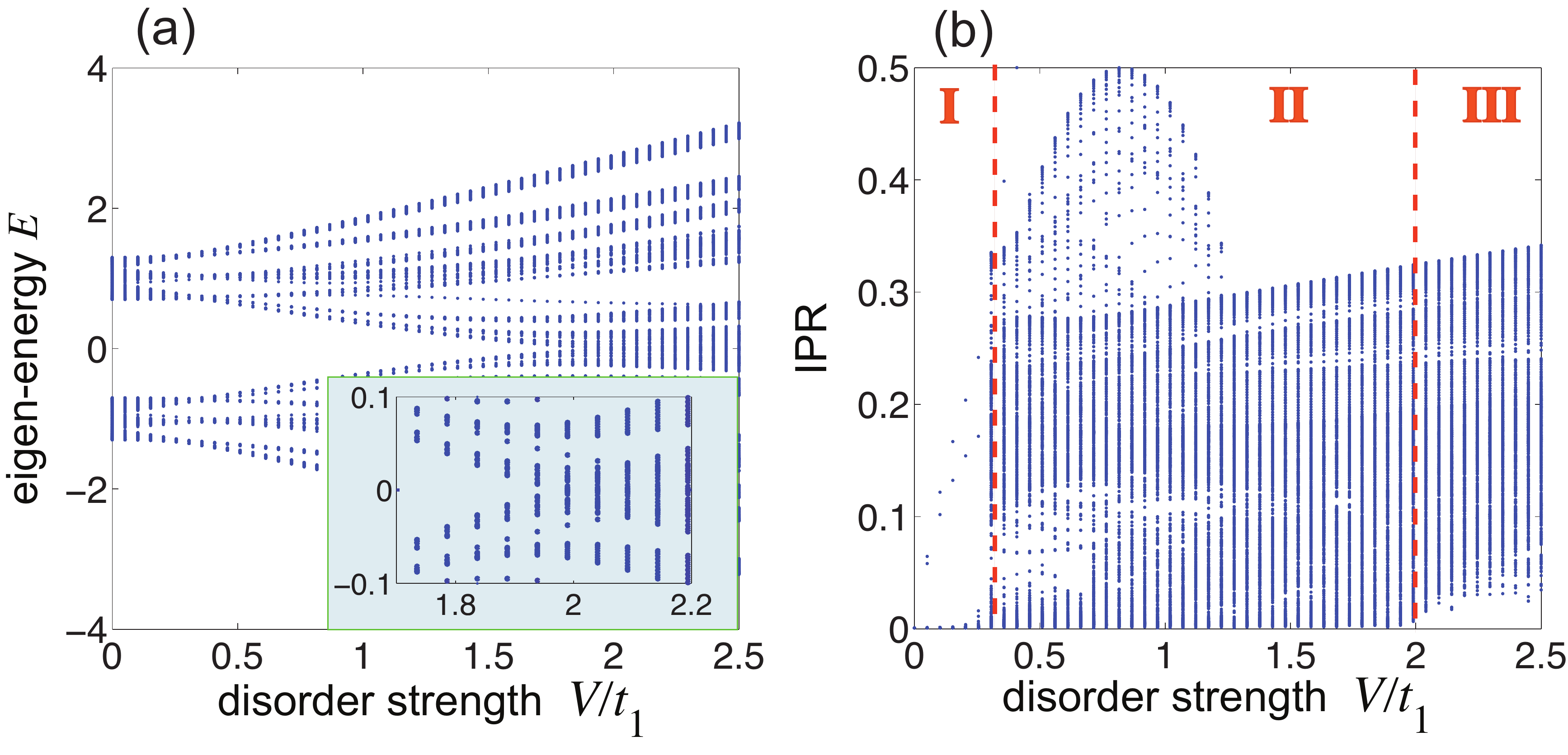}} \caption{
(Color online) Numerically-computed behavior of (a) energy spectrum $E$, and (b) corresponding IPR of the eigenmodes versus disorder strength $V/t_1$ in a SSH lattice with $t_2=0.3t_1$ comprising $N=987$ unit cells. OBC are assumed. The inset in (a) shows an enlargement of the energy spectrum near the gap closing point $V=2t_1$ ( topological Anderson phase transition). For $V>2t_1$  two nearly-degenerate zero-energy modes, localized at the left and right edges of the chain, are found. In (b), three distinct regions are clearly observed: Region I ($V<t_2$) with almost all eigenstates delocalized in the lattice (IPR $\simeq 0$); Region II ($t_2<V<2t_1$), where almost all eigenstates become localized; Region III ($V>2 t_1$), where all eigenstates are localized. The boundary between regions II and III corresponds to the gap closing point in (a).}
\end{figure} 
{\it  Topological Anderson transition and zero-energy edge states.} Here we are mostly interested in the phase transition between regions II and III at $V=2 t_1$, corresponding to band gap closing and to the appearance of zero-energy eigenmodes [Fig.2(a)]. This phase transition is precisely related to the  topological Anderson phase. In fact, for a disordered SSH chain with chiral symmetry the topological number $Q$ --that generalizes the winding number of the disorder-free SSH chain-- reads \cite{r32}
\begin{equation}
Q=\frac{1}{2}(1-Q^{\prime})
\end{equation}
where 
\begin{equation}
Q^{\prime}={\rm sign} \left\{ \prod_n t_1^2 - \prod_n (t_2+V_n)^2 \right\}.
\end{equation}
To show that the phase transition at $V=2t_1$ between regions II and III corresponds to a change of the topological number $Q$, let us notice that in the large $N$ limit one can write $Q^{\prime}={\rm sign} \{ t_1^{2N}-S^{2N} \}$, where $\log S=\lim_{N \rightarrow \infty} (1/N) \sum_{n=1}^{N} \log |t_2+V \cos (2 \pi \alpha n)|$. From Weyl's equidistribution
theorem, for $\alpha$ irrational the quantity $k = 2\pi \alpha n$  mod $2 \pi$  uniformly fills the interval $(-\pi, \pi)$ as the integer $n$ varies, so that \cite{r33}
\begin{equation} 
\lim_{N \rightarrow \infty} \frac{1}{N} \sum_{n=1}^{N} \log |t_2+V_n| = \frac{1}{2 \pi} \int_{-\pi}^{\pi} dk \log |t_2+V \cos k|.
\end{equation}
Therefore, $Q^{\prime}=1$ for $U<0$ and $Q^{\prime}=-1$ for $U>0$, where 
\begin{equation}
U \equiv \frac{1}{2 \pi} \int_{-\pi}^{\pi} dk  \log \left| \frac{t_2+ V \cos k}{t_1} \right|.
\end{equation}
The integral on the right hand side of Eq.(6) can be calculated in a closed form \cite{r33}, yielding $U>0$ for  $V>2 t_1$ and $U<0$ for  $V<2 t_1$. This means that $Q^{\prime}=1$ (and thus $Q=0$) for $V<2t_1$, and $Q^{\prime}=-1$ (and thus $Q=1$) for $V>2t_1$, corresponding to the  topological Anderson phase.  It can be readily proven that in the  topological Anderson phase $Q=1$ the SSH lattice sustains zero-energy edge states. The zero-energy mode can be exactly computed; for the semi-infinite chain with the left edge, it is defined by the recurrence relations $b_n=0$, $a_{n+1}=-t_1/(t_2+V_n) a_n$, which follow from Eqs.(1) and (2) after setting $da_n/dz=db_n/dz=0$. This state is a localized edge state provided that the Lyapunov exponent
\begin{equation}
\mu= - \lim_{n \rightarrow \infty} \frac{1}{n} \log \left| \frac{a_{n+1}}{a_1} \right|
\end{equation}
 \begin{figure*}[htb]
\centerline{\includegraphics[width=18.4cm]{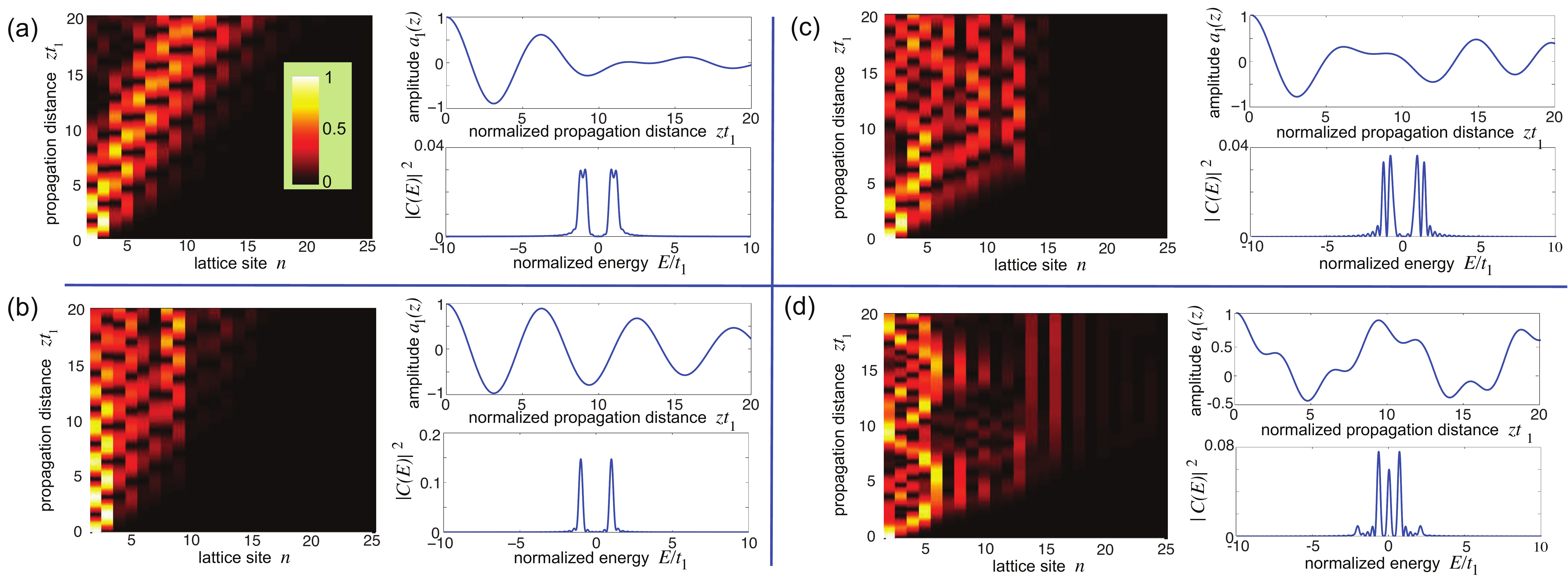}} \caption{ 
(Color online) Edge light dynamics in a SSH waveguide lattice for $t_2/t_1=0.3$ and for increasing values of the disorder strength $V$: (a) $V=0$ (disorder-free lattice), (b) $V=0.2 t_1$ (region I), (c) $V=t_1$ (region II), and (d) $V=2.5 t_1$ (region III). The sample length is $L=20/ t_1$. In each plot, the left panel shows on a pseudo color map the evolution of light intensity distribution ($|a_n|^2$ in odd sites, $|b_n|^2$   in even sites); the top right panel shows the behavior of the field amplitude $a_1(z)$ in the left-edge waveguide versus propagation distance $z$; the bottom right panel shows the  behavior of the correlation function $|C(E)|^2$.}
\end{figure*} 
is strictly positive, while for $\mu<0$ the zero energy does not belong to the energy sectrum. Since $|a_{n+1}/a_1|= \prod_{l=1}^n |t_1/(t_2+V_l)|$, one has
\begin{equation}
\mu=  \lim_{n \rightarrow \infty} \frac{1}{n} \sum_{l=1}^{n} \log \left| \frac{t_2+V_l}{t_1} \right|= \frac{1}{2 \pi} \int_{-\pi}^{\pi} dk \log \left| \frac{t_2+V \cos k }{t_1} \right|
\end{equation}
i.e. one has $\mu=U$, where $U$ is defined by Eq.(6). Hence, the zero-energy edge state does exist when $U>0$, i.e. when the topological number is $Q=1$, while it does not exist when $U<0$, i.e. when the topological number is $Q=0$.\\
\\
{\it Edge dynamics and  topological Anderson phase.} A simple method to probe the topological phase of the disorder-free SSH lattice is to exploit the bulk-boundary correspondence, i.e. the existence of a localized edge state in the topological phase $Q=1$. This entails to excite the lattice at the edge site at input plane $z=0$ and to check whether some light remains trapped at the edge \cite{r30}. Unfortunately, this method can not be extended to a lattice with disorder because modes become rapidly localized at disorder strengths smaller than the critical value $V=2t_1$ [Fig.2(b)]. Figure 3 shows a few examples of light dynamics in disordered lattices at a few increasing values of $V/t_1$ for $t_2=0.3 t_1$ and for single waveguide excitation of the left site. As expected, in most cases some light remains trapped at the left edge, however this is not a signature of the  topological Anderson phase. While bulk methods, such as the mean chiral displacement method \cite{r34,r34b}, can be used to detect  topological Anderson transition in the disordered SSH model  \cite{r21}, an open question is whether edge light dynamics could provide a clear signature of phase transition. Here we show that, despite localization induced by disorder, the use of spectral methods \cite{r27,r35} can provide a signature of the  topological Anderson phase when considering light dynamics at the edge of the lattice. To this aim, let $a_1(z)$ be the evolution of the light field in the left-edge waveguide of the lattice, when it is excited at the input plane [$a_n(0)=\delta_{n,1}$, $b_n(0)=0$], and let us consider the correlation function 
\begin{equation}
C(E)=\frac{1}{L}\int_0^L dz \; a_1^*(0)a_1(z) \exp(iEz)
\end{equation}
 where $L$ is the propagation length in the sample. Note that $C(E)$ reduces to the sampled Fourier transform of $a_1(z)$ for the single-site input excitation $a_1(0)=1$.
Indicating by $\alpha_l$ the spectral weight of the initial lattice excitation into the $l$-th eigenmode with eigenenergy $E_l$, the correlation function $C(E)$ takes the form \cite{r27,r34}
 \begin{equation}
 C(E)=\sum_l \alpha_l G(E-E_l)
 \end{equation}
where $|G(E)|^2=[ \sin (EL/2)]^2/ (EL/2)^2$. Note that $G(E)$ is a narrow function, i.e. $|G(E)|^2 \sim 2 \pi \delta(E)$, with a resolution $ \Delta E \sim 2 \pi /L$ which is determined by the maximum propagation length $L$ available in an experiment. From the behavior of $|C(E)|^2$ one can thus get some information about the energy spectrum of the lattice. In an experiment, the correlation function $|C(E)|^2$ can be readily reconstructed from light intensity measurements following a simple four-step procedure, as discussed in a recent experiment \cite{r27}: Excite the left-edge waveguide of the array; monitor the intensity light evolution $I(z)=|a_1(z)|^2$ along $z$ (which is routinely done in fs laser written waveguides by fluorescence imaging \cite{r3});
retrieve the evolution of the $a_1(z)$ of the light mode by taking the square root of $I(z)$, with  an appropriate choice of the sign;
numerical computation of the correlation function from the experimental trace $a_1(z)$.
 From Fig.2(a), it is clear that for $V<2 t_1$ the gap is open and there is not the zero-energy left edge mode: in this case $|C(E)|^2$ does not show any peak centered at around $E=0$, as illustrated in Figs.3(a) and (b) and (c). As $V$ crosses the phase transition point $2t_1$, a peak at around $E=0$ is instead observed, as shown in Fig.3(d). To highlight the suitability of the spectral method to detect the topological phase transition, in Fig.4 we plot the behavior of the integrated spectral density $F=\int_{- \Delta E}^{\Delta E} dE |C(E)|^2$, versus $V/t_1$, for increasing values of the sample length $L$. The figure clearly shows that $F$ is almost vanishing for $V<2t_1$, while it increases almost linearly with $V$ above the phase transition point $V=2t_1$. Note that the resolution of the transition is sharper as a longer propagation distance $L$ is available.\\ 
The results presented in this work are feasible for an experimental observation using waveguide arrays manufactures by fs laser writing \cite{r3,r27}. 
The sequence of alternating coupling constants $t_1$ (intra-dimer hopping) and $t_2+V_n$ (inter-dimer hopping) can be realized with precise accuracy by a judicious spacing $d$ between adjacent waveguides \cite{r26,r27}, taking into account the near exponential dependence of the coupling constant on $d$.
For example, let us assume a waveguide lattice in fused silica probed with red light, and let us assume the following dependence of coupling constant $t$ on waveguide spacing $d$:  $t \simeq t_0 \exp[-\gamma (d-a)]$, with $t_0=1.27 \; {\rm cm}^{-1}$, $a= 15 \; \mu$m and $\gamma=0.20 \; {\mu}^{-1}$ (such parameters are taken from the experimental data of Ref.\cite{r27}). 
To avoid the sign change of $t_2+V_n$ for $V>t_2$, which is not experimentally feasible, one can take $|t_2+V_n|$ rather than $t_2+V_n$ in the waveguide array design; this is basically equivalent to  introduce  at some sites a sign change of $b_n$, however the light dynamics at the edge is not modified. For an intra-dimer homogenous spacing $a=15 \; \mu$m, one has $t_1=t_0$, and for the largest value of disorder strength $V/t_1=2.5$ considered in the simulations of Figs.3 and 4, the largest value of $|t_2+V_n|$ is $2.8 t_1 \simeq 3.56 \; {\rm cm}^{-1}$, corresponding to a minimum waveguide spacing of $\sim10 \; \mu$m. For a maximum propagation length $L=15 / t_1$ (green curve in Fig.4), a sample length of $L \simeq 11.8$ cm is required, which is feasible with current fs laser writing technology.

 \begin{figure}[htb]
\centerline{\includegraphics[width=8.4cm]{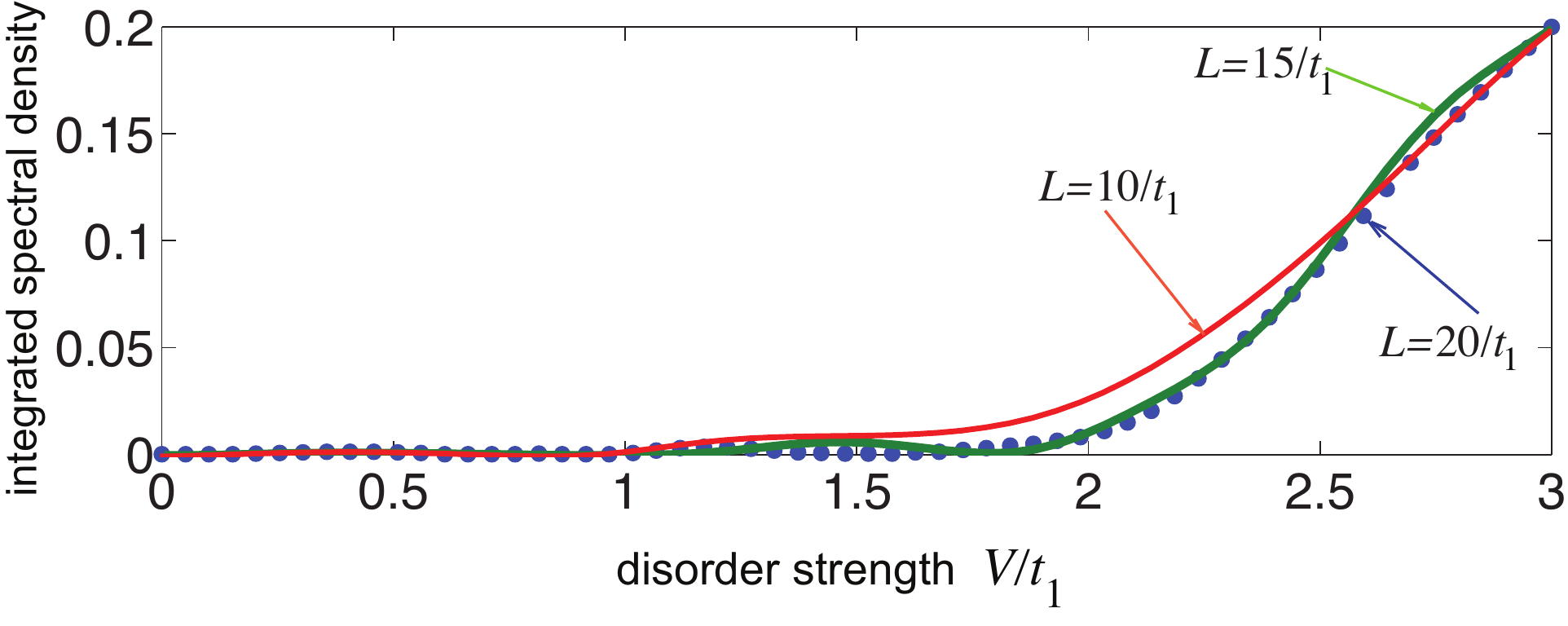}} \caption{ 
(Color online) Numerically-computed behavior of the integrated spectral density $F$ versus disorder strength $V$ for a few increasing values of the maximum propagation distance $L$.
Other parameter values are as in Fig.3.}
\end{figure}

{\it Conclusions.} We predicted a  topological Anderson phase in a SSH waveguide lattice where inter-dimer disorder is quasi-periodic (rather than random). This phase is characterized by a non-vanishing topological number and by the existence of zero-energy edge modes, which can be detected by a spectral method analysis from edge intensity light dynamics. Our results, besides of providing major physical insights into the interplay between correlated disorder and Anderson topological pheses, could be also of interest for the design of a new class of topological lasers. In topological lasers based on active SSH lattices so far reported \cite{r36,r37,r38}, the SSH chain is in a topological nontrivial phase and lasing in a topologically-protected edge state is obtained by selective pumping. The addition of small-to-moderate disorder in the coupling costants does not change the emission frequency of the lasing mode, but strong disorder fully destroys topological protection. The topological Anderson phase transition suggests that a different type of topological laser based on an active SSH chain can be realized: lasing in a topologically-protected edge state by selective pumping is observed under strong disorder conditions of inter-dimer coupling constants, albeit averaging over disorder the SSH lattice remains in a trivial topological phase.\\
The author declares no conflicts of interest.

\end{document}